\documentclass{article}
\usepackage{ijcai20}

\usepackage{times}
\usepackage{soul}
\usepackage{url}
\usepackage[utf8]{inputenc}
\usepackage[small]{caption}
\usepackage{graphicx}
\usepackage{amsmath}
\usepackage{amsthm}
\usepackage{booktabs}
\usepackage{algorithm}
\usepackage{algorithmic}

\usepackage{subfigure}
\usepackage{amsfonts}
\usepackage{multirow}
\newtheorem{theorem}{Theorem}
\newtheorem{definition}{Definition}

\title{EPINE: Enhanced Proximity Information Network Embedding}

\author{
    Luoyi Zhang, Ming Xu
    \affiliations
    State Key Laboratory for Novel Software Technology, Nanjing University, Nanjing, China
    \emails
    zhangluoyi.cs@gmail.com, xuming0830@gmail.com
}

\begin{document}

\maketitle

\begin{abstract}
    Unsupervised homogeneous \textit{network embedding} (NE) represents every vertex of networks into a low-dimensional vector and meanwhile preserves the network information. Adjacency matrices retain most of the network information, and directly charactrize the first-order proximity. In this work, we devote to mining valuable information in adjacency matrices at a deeper level. Under the same objective, many NE methods calculate high-order proximity by the powers of adjacency matrices, which is not accurate and well-designed enough. Instead, we propose to redefine high-order proximity in a more intuitive manner. Besides, we design a novel algorithm for calculation, which alleviates the scalability problem in the field of accurate calculation for high-order proximity. Comprehensive experiments on real-world network datasets demonstrate the effectiveness of our method in downstream machine learning tasks such as network reconstruction, link prediction and node classification. 
\end{abstract}

\section{Introduction}\label{sec1}
We live in a world where many things form network structures, including social networks, citation networks and word cooccurrence networks. Under the circumstances, network embedding is proposed to process and exploit the network data, which facilitates downstream machine learning tasks such as network reconstruction, link prediction and node classification.

Most information contained in ubiquitous networks is reflected by the proximity between nodes, or weights of edges, which are stored in adjacency matrices. Hence NE is to map every network vertex into a low-dimensional vector that preserves adjacency matrix information as much as possible. If there is some similarity between two objects, NE would represents them by similar representations, such as vectors that have low Euclidean distance \cite{luo2011cauchy}, or certain dot product values \cite{tang2015line,ou2016asymmetric,liu2019general}. Meanwhile, to emphasize the similarity, methods such as negative sampling are proposed to dispart the representations of objects that seem to be different. In a word, similar things would possess similar representations, and vice versa \cite{liu2019general}. In this work, network vertex embeddings are learned in two steps: first defining and calculating the similarity, then preserving it into the node embeddings.

The core problem we focus on is how to search and define the similarity between nodes. First-order proximity acts as a straightforward approach that merely considers two endpoints of an edge to be similar, and was researched by previous graph embedding methods \cite{belkin2003laplacian}. To supplement this, higher-order proximity was proposed. 
LINE \cite{tang2015line} has defined second-order proximity as relationships that there exist common neightbors between two unconnected nodes. Noting that common neighbors are also the intermediate vertices of length-two shortest paths. Intuitively, $k$-order proximity could be analogously defined as the relations that there are shortest paths of length $k$ between nodes, and higher-order indicates a weaker relationship.

\begin{figure*}[tbp]
\centering
\subfigure[Samples of 2-step walks]{
\label{fig2.sub.1}
\includegraphics[height=1.in]{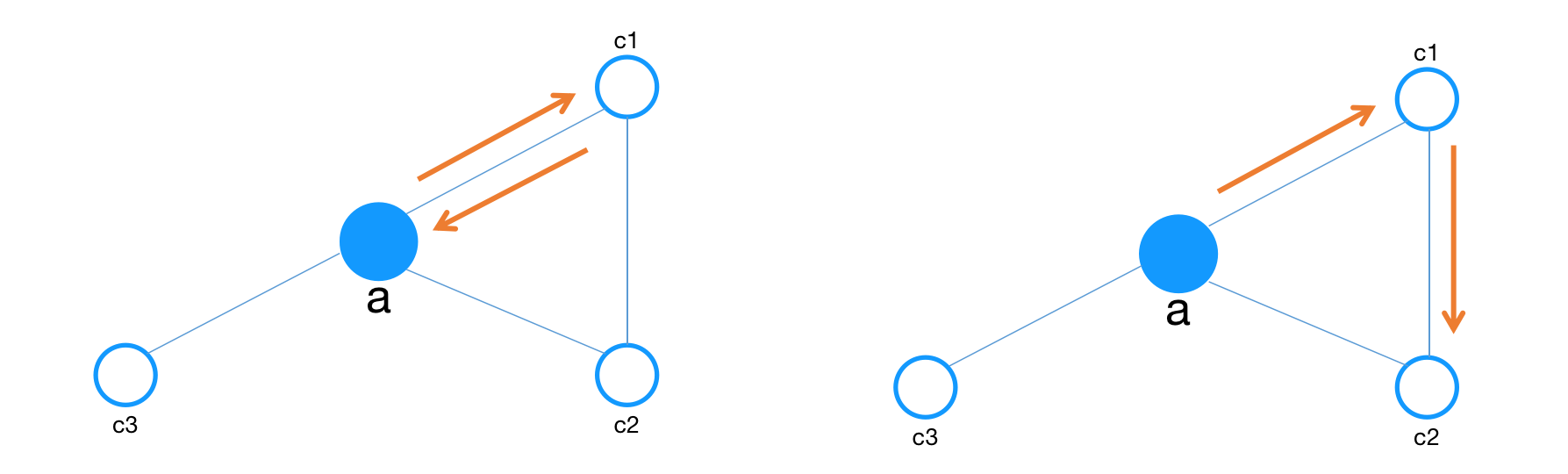}}
\subfigure[Samples of 3-step walks]{
\label{fig2.sub.2}
\includegraphics[height=1.in]{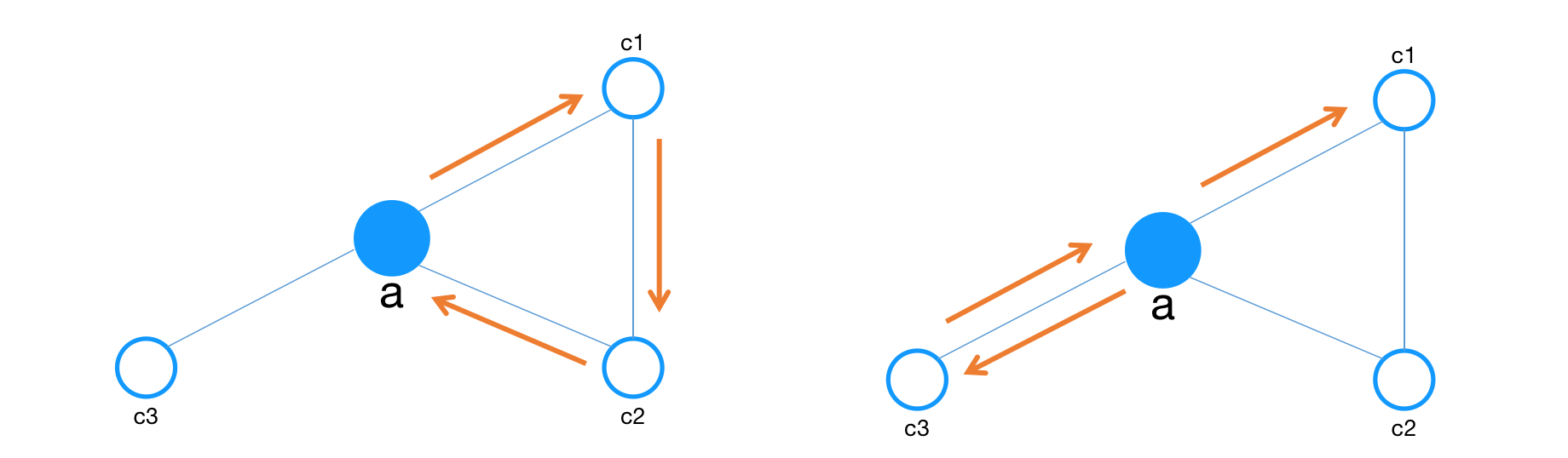}}

\caption{A simplified illustration of $k$-step walks starting from node $v_a$.}
\label{fig2}
\end{figure*}

However, many previous network embedding methods \cite{cao2015grarep,perozzi2017don,zhang2018arbitrary} proposed to treat the $k^{th}$ power of adjacency matrices or probability transition matrices as $k$-order proximity matrices. It's worth noting that in unweighted networks, the values of nonzero elements in matrices state the number of $k$-step walks between vertices, which may contain repetitive edges. As shown in Figure \ref{fig2}, through a two- or three-step walk, one can return back to the start point, or reach immediate neighbor. Therefore, defining $k$-order proximity as $k$-step walks is not accurate and well-designed enough. Actually, the $k^{th}$ power of adjacency matrices contains a mixture of proximity no more than $k^{th}$ order.

Consequently, we propose Enhanced Proximity Information Network Embedding (EPINE), a novel approach to redefine the high-order proximity, and determine the strength of $k$-order proximity by the number of length-$k$ shortest paths between nodes and weights along with the paths. With regard to calculation, we develope a novel algorithm that has the same complexity as the powers of adjacency matrices, which alleviates the scalability problem. Due to the close relation to edge weights in calculation, for unweighted networks, we propose to assess every edge weight by the degrees of two endpoints.

In addition, weight information carried by edges can be regarded as another metric to evaluate the node similarity. However, due to the lack of weight information stored in network datasets, we might have to treat every edge equally, which is out of line with the reality. Take social networks as an example, friends may fall into five categories: bosom friends, good friends, ordinary friends, acquaintances and prequaintances. The strength of friendships among them decrease progressively, but edges are of equal importance in most social network datasets. 

Eventually, the similarity measurement in EPINE is distance-based (and structure-based for unweighted networks, where node degrees as structural information), and we name it as EPINE similarity.

In summary, our contributions are as follows:

\begin{itemize}
\item We redefine the high-order proximity in a more accurate and intuitive manner, and propose a novel approach for calculation that alleviates the scalability problem.
\item We conduct comprehensive experiments on real-world network datasets. Experimental results demonstrate the effectiveness of the proposed EPINE. 
\end{itemize}

The rest of the paper is arranged as follows. In Section \ref{sec2}, we introduce some preliminaries and discuss related work. In Section \ref{sec3}, we depict our proposed method in detail. We outline experimental results and analyses in Section \ref{sec4}, and close with conclusions and future work in Section \ref{sec5}.

\section{Preliminaries and Related Work}\label{sec2}

\subsection{Notations and Definitions}

A network is denoted as $\mathcal{G}=(\mathcal{V}, \mathcal{E})$, where $\mathcal{V}=\{v_i|i=1,2,\cdots,|\mathcal{V}|\}$ is the node set and $\mathcal{E}$ is the edge set. $e_{ij}$ in $\mathcal{E}$ has a binary value that indicates the existence of an edge from node $i$ to node $j$. The weight $w_{ij}$ is equal to $e_{ij}$ in unweighted networks and a non-negative value in weighted ones.

The adjacency matrix $A \in \mathbb{R}^{|\mathcal{V}|\times|\mathcal{V}|}$ is defined as $A_{i,j} = w_{ij}$. $d_i=\sum_j A_{i,j}$ is the degree of node $i$, and the diagonal degree matrix $D \in \mathbb{R}^{|\mathcal{V}|\times|\mathcal{V}|}$ has the element $D_{i,i} = d_i$. The (one-step) probability transition matrix, also called normalized adjacency matrix, is obtained by $\hat{A} = D^{-1}A$.

In graph theory, a \textit{walk} consists of an alternating sequence of vertices and edges that begins and ends with a vertex. A \textit{path} is a walk without repeated vertices.

We differentiate the general $k$-order ($k \in \mathbb{N^+}$) proximity formally defined in \cite{zhang2018network} and our redefined one in the following.

\begin{definition}\label{def1}
(Vanilla $k$-order Proximity). It would be a vanilla $k$-order proximity relationship between two nodes if and only if there exists at least one walk of length $k$ between them.
\end{definition}

We depict an unweighted and undirected ego network in Figure \ref{fig2}, and denote the $k^{th}$ power of adjacency matrix as $A^k$. Note that node $c_{1,2,3}$ are all immediate neighbors of node $v_a$, but $A^2_{a,a}=3$, $A^2_{a,c_1}=A^2_{a,c_2}=1$, $A^3_{a,a}=2$, and $A^3_{a,c_1}=A^3_{a,c_2}=A^3_{a,c_3}=3$. Actually, each nonzero element of $A^k$ denotes a vanilla $k$-order proximity, which is shown to be not intuitive and accurate enough. Consequently, we proposed to redefine it as follow:

\begin{definition}\label{def2}
(Rectified $k$-order Proximity). Two nodes have a rectified $k$-order proximity relationship if and only if there exists at least one shortest path of length $k$ between them.
\end{definition}

Accordingly, we denote the rectified $k$-order proximity matrix as $A^{k-order} \in \mathbb{R}^{|\mathcal{V}|\times|\mathcal{V}|}$, where positive elements represent rectified $k$-order proximity between nodes.

In a sense, vanilla proximity is an approximate form of rectified proximity, because every shortest path of length $k$ is also a $k$-step walk, which also consists in $A^k$.

\subsection{Related Work}

\subsubsection{Preserving High-order Proximity}\label{sec2.3.1}

Almost every network embedding method would preserve first-order proximity, while higher-order proximity acts as complementary and global information of networks, and is explored by a bunch of methods.

\textbf{DeepWalk} \cite{perozzi2014deepwalk} implicitly preserved proximity no more than $t^{th}$ order, where $t$ is the window size, and lower orders have higher weights. It can also be interpreted as factorizing a matrix $M$ \cite{yang2015network}, where

\begin{equation}\label{eq1}
  M = \log \frac{\hat{A}+\hat{A}^2+\cdots+\hat{A}^t}{t}.
\end{equation}
There are extensions and improvements of DeepWalk. \textbf{Node2vec} \cite{grover2016node2vec} substituted random walks by breadth-first and depth-first walks. \textbf{\textsc{Walklets}} \cite{perozzi2017don} replaced the adjacency matrix $A$ used in DeepWalk by one or more different powers of $A$.
\textbf{GraRep} \cite{cao2015grarep} obtained vertex embeddings by separately factorizing $\{\hat{A},\hat{A}^2,\cdots,\hat{A}^t\}$ and concatenating the results at last. 
\textbf{HOPE} \cite{ou2016asymmetric} constructed a framework that built node embeddings according to high-order proximity measurements, including Katz Index, Rooted PageRank, Common Neighbors and Adamic-Adar, which are all determined by matrix-chain multiplications of $A$ or $\hat{A}$.
\textbf{AROPE} \cite{zhang2018arbitrary} proposed to exploit arbitrary-order proximity by factorizing a matrix $S$, where

\begin{equation}\label{eq2}
  S = w_1A + w_2 A^2 + \cdots + w_k A^k,
\end{equation}
and allow $k=+\infty$ if the summation converges.

Algorithms mentioned above derive high-order proximity more or less from the power of $A$ or $\hat{A}$, which is actually the vanilla high-order proximity.

Besides, \textbf{SDNE} \cite{wang2016structural} preserved rectified second-order proximity by reconstructing adjacency matrices, and retained first-order proximity by a regularization term derived from laplacian eigenmaps \cite{belkin2003laplacian}. \textbf{LINE} \cite{tang2015line} proposed to represent and restore first-order and second-order proximity in content and context representations respectively. Compared with our proposed method, they can only calculate fixed-order proximity, rather than arbitrary-order proximity.

\subsubsection{Capturing Edge Information}

Existing NE methods that consider edge information \cite{tu2017transnet,goyal2018capturing,chen2018enhanced} generally rely on the intrinsic edge information stored in network datasets. Instead, we derive edge weights from node degrees, which is independent of intrinsic edge information, and applicable to multifarious networks.

\section{Enhanced Proximity Information Network Embedding}\label{sec3}

In this section, we describe how to calculate accurate $k$-order proximity matrix $A^{k-order}$ in detail.

Suppose ${\gamma}^k_{ij}=(v_{i_0},e_{i_0,i_1},v_{i_1},\cdots,v_{i_k}) $ where $i_0=i, i_k=j$ is a length-$k$ walk between $v_i$ and $v_j$, then 

\begin{equation}\label{eq8}
  A^k_{i,j} = \sum_{\gamma^k_{ij}} f_{cost}(\gamma^k_{ij}),
\end{equation}
where

\begin{equation}\label{eq9}
  f_{cost}(\gamma^k_{ij}) = w_{i_0,i_1}\cdot w_{i_1,i_2} \cdots w_{i_{k-1},i_k}
\end{equation}
is the chain multiplication of edge weights along the walk $\gamma^k_{ij}$.

In unweighted networks, every edge weight is set to 1, that is, $f_{cost}(\gamma^k_{ij})=1$ and $A^k_{i,j}$ states the number of $k$-step walks between $v_i$ and $v_j$.

We discover that $k$-step walks contain all $k$-length shortest paths, hence we can extract the latter through a second-order deterministic process.

\subsection{Calculating the Rectified Proximity}

For the convenience of statement, we first define the $k$-reachable relationship.

\begin{definition}\label{def3}
  ($k-reachable$). Node $v_j$ is $k$-reachable from $v_i$ if and only if there exists a shortest path of length $k$ from $v_i$ to $v_j$.
\end{definition}

In the $i$-th row of $A^{k-order}$, each positive element stands for a $k$-reachable node of $v_i$. Then we can calculate the rectified $k$-order proximity based on the following theorem:

\begin{theorem}
  The matrix product $A^{k-order} \cdot A$ contains and only contains rectified proximity of order (k-1), k and (k+1). Such (k+1)-order proximity composes the $A^{(k+1)-order}$.
\end{theorem}

\begin{figure}[tbp]
\centering
\subfigure[One-step walk]{
\label{fig3.sub.1}
\includegraphics[height=1.3in]{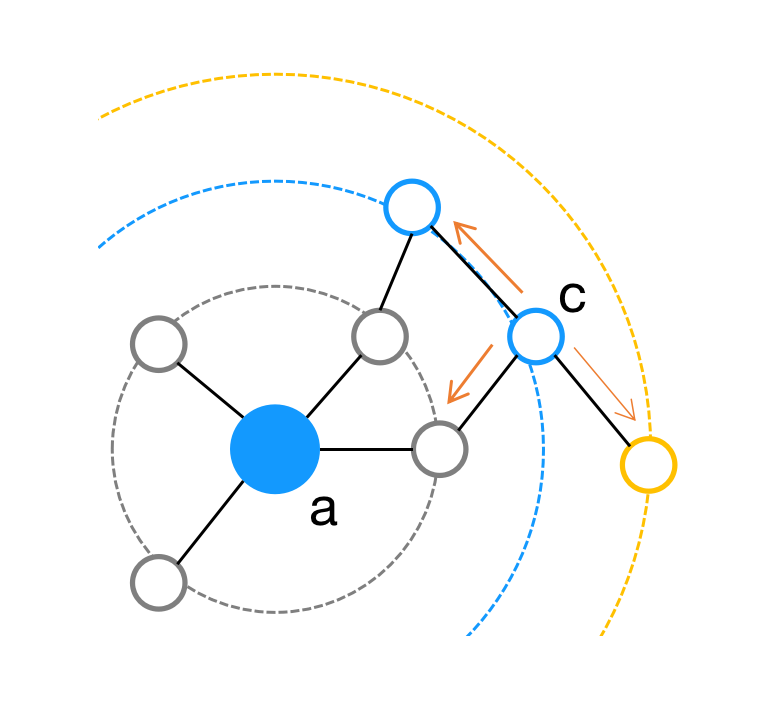}}
\subfigure[Masking]{
\label{fig3.sub.2}
\includegraphics[height=1.3in]{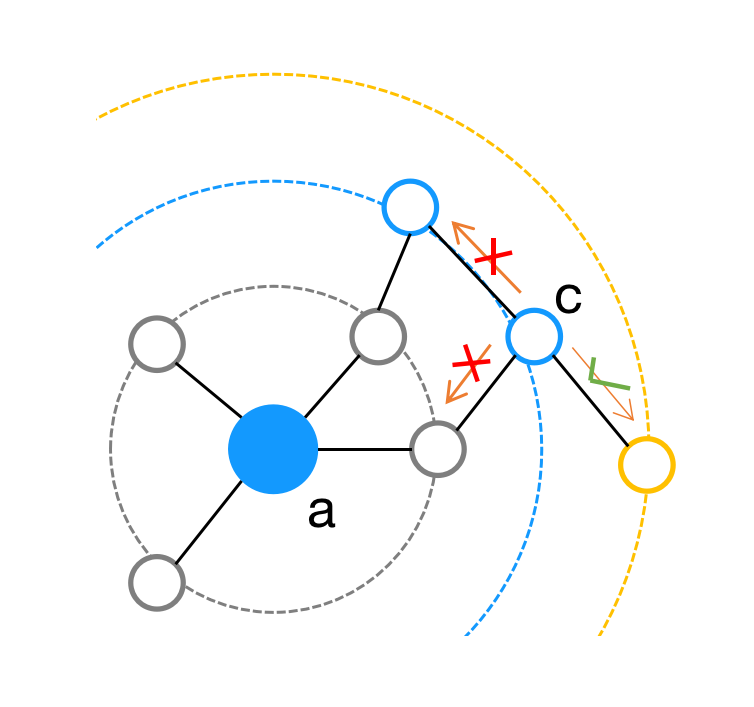}}

\caption{A simplified illustration.}
\label{fig3}
\end{figure}

\noindent \textbf{Proof.} Post-multiplying $A^{k-order}$ by $A$ actually performs one-step walks starting from every node. As illustrated in Figure \ref{fig3.sub.1}, the grey, blue and yellow circle denote the reachability of $(k-1)$, $k$ and $(k+1)$, respectively. Suppose node $v_c$ is $k$-reachable from node $v_a$. Moving one step from node $v_c$ would only result in three categories of situations: moving backwards to the grey circle, staying in the blue, or walking farther to the yellow, which all generate $(k+1)$-length walks and are actually rectified proximity of order $(k-1)$, $k$ and $(k+1)$ respectively. 

On the other hand, any $(k+1)$-length shortest path consists of a $k$-length shortest path and another one edge, hence walking farther from the blue circle to the yellow forms all $(k+1)$-length shortest paths that compose $A^{(k+1)-order}$. \hfill $\square$

Consequently, we can extract $(k+1)$-order proximity by removing rectified proximity of order $(k-1)$ and $k$ from $A^{k-order} \cdot A$. 

By Definition \ref{def2}, any two rectified proximity matrices are mutually disjoint, which means if $A^{(k-1)-order}_{i,j}$ or $A^{k-order}_{i,j}$ is nonzero, then $A^{(k+1)-order}_{i,j}=0$. That is, if we have $A^{(k-1)-order}$ and $A^{k-order}$, we could accurately calculate $A^{(k+1)-order}$ via applying masking to $A^{k-order} \cdot A$, which can be calculated by

\begin{equation}\label{eq11}
  Mask^{k+1}_{i,j} = \begin{cases}
    0, \quad if\ (A^{(k-1)-order} + A^{k-order})_{i,j} > 0;\\
    1, \quad \text{otherwise}.
  \end{cases}
\end{equation}

Such a calculating method could be regarded as a second-order deterministic process, which is described in Algorithm \ref{alg1}. As stated in line 1, we extend the Definition \ref{def2} and define $A^{0-order}$ as $I_{|\mathcal{V}|}$, which treats self-loops as rectified $0$-order proximity. Then we can obtain $A^{k-order}$ through

\begin{equation}
\begin{aligned}
  A^{0-order} &= I_{|\mathcal{V}|},\ A^{1-order} = A,\\
  A^{k-order} &= f_{mm}(A^{(k-1)-order}, A)\ \odot \\
  &\quad\ f_{mask}(A^{(k-2)-order}, A^{(k-1)-order}),
\end{aligned}
\end{equation}
where $f_{mm}(\cdot,\cdot)$ represents matrix multiplication (discussed later in Section \ref{sec3.3}), $\odot$ means Hadamard product, and $f_{mask}(\cdot,\cdot)$ calculates the mask based on Equation (\ref{eq11}). It's worth noting that in practice, the above-mentioned $k$ always has a finite value, which depends on the longest shortest paths in networks. Consequently, the algorithm should early stop if and only if $A^{new}$ in line 8 is a zero matrix.

\begin{algorithm}[tbp]
\renewcommand{\algorithmicrequire}{\textbf{Input:}}
\renewcommand{\algorithmicensure}{\textbf{Output:}}
\caption{Calculating Rectified $k$-order Proximity}
\label{alg1}
\begin{algorithmic}[1]
    \REQUIRE Weighted Adjacency Matrix $A$, Order Number $k(k\geq 2)$, Matrix Multiplication Function $f_{mm}(\cdot,\cdot)$, Mask Calculation Function $f_{mask}(\cdot,\cdot)$
    \ENSURE Rectified $l$-order proximity matrix $A^{l-order}(l\leq k)$
    \STATE Initialize: $A^{last} \gets I_{|\mathcal{V}|}$ 
    \STATE Initialize: $A^{current} \gets A$
    \STATE Initialize: $A^{new} \gets $ None
    \FOR {$l=2 \to k$}
    \STATE $A^{new} \gets f_{mm}(A^{current}, A)$
    \STATE $Mask^l \gets f_{mask}(A^{last}, A^{current})$
    \STATE $A^{new} \gets A^{new} \odot Mask^l$
    \IF{$A^{new}=\mathbf{0}$}
    \STATE break
    \ENDIF
    \STATE $A^{last} \gets A^{current}$
    \STATE $A^{current} \gets A^{new}$
    \ENDFOR
    \RETURN $A^{current}$
\end{algorithmic}
\end{algorithm}

\subsection{Rethinking the Matrix Multiplication}\label{sec3.3}

If we apply normal matrix multiplication as the implementation of the function $f_{mm}(\cdot,\cdot)$, similar to Equation (\ref{eq8}) and (\ref{eq9}), the weight (or cost) of a path $p^k_{ij}$ would be

\begin{equation}\label{eq13}
  f_{cost}(p^k_{ij}) = w_{i_0,i_1}\cdot w_{i_1,i_2} \cdots w_{i_{k-1},i_k},
\end{equation}
then

\begin{equation}
  A^{k-order}_{i,j} = \sum_{p^k_{ij}} f_{cost}(p^k_{ij}).
\end{equation}

Note that the cost is obtained by the chain multiplication of edge weights along the path. By definition, edge weights are always larger than 1, which would cause very large path costs. Such large path costs might not be discriminative and effective enough in practice.

Among the sum, mean and max aggregators, sum has the best discriminative power \cite{xu2019powerfulgnns}. Hence we resort to additive operation, that is, we could calculate path costs through 

\begin{equation}\label{eq15}
  f_{cost}(p^k_{ij}) = w_{i_0,i_1} + w_{i_1,i_2} + \cdots + w_{i_{k-1},i_k},
\end{equation}
which also alleviates the explosion of path costs. For the sake of calculating Equation (\ref{eq15}), we propose to adopt \textit{additive matrix multiplication} to $f(\cdot,\cdot)$, which could be formalized as (suppose $f(X,Y)=Z$):

\begin{equation}
  Z_{i,j} = \sum_{t=1}^r \mathbb{I}(X_{i,t} \cdot Y_{t,j}) \cdot (X_{i,t} + Y_{t,j}),
\end{equation}
where $r$ is the column and row number of the matrix $X$ and $Y$ respectively. $\mathbb{I}(\cdot)$ is a indicator function that 

\begin{equation}
  \mathbb{I}(x) = \begin{cases}
    1, \quad x\neq 0;\\
    0, \quad \text{otherwise}.
  \end{cases}
\end{equation}

\subsection{Calculating Edge Weights}

Due to the close relation to edge weights in Equation (\ref{eq15}), we calculate edge weights as inputs to Algorithm \ref{alg1} for unweighted networks. Intuitively, edges connected to low-degree nodes would be more decisive, which is in accordance with the degree penalty principle \cite{feng2018scalefree}. Hence we could evaluate edge weights simply by

\begin{equation}
  w_{ij}=\frac{1}{d_i \cdot d_j}.
\end{equation}

\subsection{The Proposed EPINE Similarity}

Eventually, we define the \textit{EPINE similarity} matrix $S^{EPINE} \in \mathbb{R}^{|\mathcal{V}|\times|\mathcal{V}|}$ as

\begin{equation}\label{eq17}
  S^{EPINE} = A^{1-order} + \alpha_2 A^{2-order}  + \cdots + \alpha_k A^{k-order}
\end{equation}

As discussed in Section \ref{sec1}, higher-order proximity indicates a weaker relationship, which is in line with the exponentially decaying weights of Katz similarity \cite{katz1953new}. Hence we propose to decay weights of rectified high-order proximity by

\begin{equation}\label{eq18}
  \alpha_i = \frac{\lambda_i}{f_{max}(A^{i-order})}(2\leq i\leq k),
\end{equation}
where $\lambda_i$ is the decay coefficient and $f_{max}(\cdot)$ returns the maximum element of the input matrix. 

As researched by previous work \cite{perozzi2014deepwalk,feng2018scalefree}, the degree distribution of networks probably follows the power law. It suggests there might be only a few elements of $A^{k-order}$ have overlarge values, and divided by them in Equation (\ref{eq18}) would degenerate the information carried by $A^{k-order}$. Consequently, before calculating Equation (\ref{eq17}), we truncate the largest $\eta\times|\mathcal{V}|^2$ elements of $A^{k-order}$ to the value of the element right smaller than them.

\subsection{Learning the Network Embedding}

As $S^{EPINE}$ could be regarded as a weighted adjacency matrix, we apply LINE \cite{tang2015line} --- a scalable method suitable for undirected, directed, and/or weighted networks --- to preserving this similarity information into vertex embeddings. To be specific, we input the EPINE similarity matrix into the LINE(1st) and LINE(2nd) so as to learn node embeddings.

\subsection{Discussions}

\textbf{Complexity.} In consideration of efficiency problems, we adopt sparse implementation for EPINE. In Algorithm \ref{alg1}, matrix multiplication has a time complexity of $O(|\mathcal{V}|d^2)$, where $d$ is the average node degree of networks. Masking step (line 6-7) takes $O(|\mathcal{E}|)$ time. The time cost for edge weight calculation and LINE are both $O(|\mathcal{E}|)$. Eventually, the overall time complexity of EPINE is $max(O(k|\mathcal{V}|d^2),\ O(|\mathcal{E}|))$, where $k$ is usually set to 2 in practice. 

\textbf{Online learning.} For any specific network, we only have to calculate $A^{k-order}$ once. When a new node $v_n$ arrives, we can calculate its similarity with existing nodes in $O(d^2)$ time, and obtain its embedding through LINE in $O(d)$ time, with embeddings of existing nodes unchanged.

\section{Experiments}\label{sec4}
In this section, we demonstrate the effectiveness of our method in three downstream machine learning tasks: network reconstruction, link prediction and node classification.  

\subsection{Datasets}

We conduct experiments on four networks. The statistics of them are listed in Table \ref{tab1}. Wikipedia is weighted, others are unweighted.

% \linespread{1.35}
\newcommand{\tabincell}[2]{\begin{tabular}{@{}#1@{}}#2\end{tabular}}
\begin{table}
\begin{center}
\caption{Dataset statistics}
\begin{tabular}{|c|c|c|c|c|}
\hline
Name & \#Nodes & \#Edges & \tabincell{c}{Avg.\\Deg.} & \#Labels\\
\hline
Wikipedia & 4,777 & 184,812 & 38.69 & 40\\
\hline
BlogCatalog & 10,312 & 333,983 & 64.78 & 39\\
\hline
Flickr & 80,513 & 5,899,882 & 146.56 & 195\\
\hline
YouTube & 1,138,499 & 2,990,443 & 5.25 & 47\\
\hline
\end{tabular}
\label{tab1}
\end{center}
\end{table}

\begin{itemize}
\item \textbf{Wikipedia} \cite{mahoney2011large}: A language network extracted from Wikipedia. The weight of each edge represents the number of co-occurrences between two words. Labels represent the Part-of-Speech (POS) tags inferred using the Stanford POS-Tagger.
\item \textbf{BlogCatalog}, \textbf{Flickr} \cite{tang2009relational}, \textbf{YouTube} \cite{tang2009scalable}: Social networks that edges indicate friendships between users, labels represent blogger interests, user groups and user groups respectively.
\end{itemize}

\subsection{Baselines and Parameter Settings}

In experiments, we compare our method with several baselines that are competitive or preserve high-order proximity. 

\begin{itemize}
  \item \textbf{DeepWalk} \cite{perozzi2014deepwalk} implicitly preserves high-order proximity and is a competitive method applicable to diverse networks.
  \item \textbf{LINE} \cite{tang2015line} contains two methods that preserves first-order and second-order proximity, which are called LINE(1st) and LINE(2nd) respectively. Besides, LINE(1st+2nd) concatenates the result embeddings of them, and LINE(rc) reconstructs networks by adding vanilla second-order neighbors to nodes' neighbors. In experiments, we report the best results of them.
  \item \textbf{GraRep} \cite{cao2015grarep} accurately calculate the vanilla high-order proximity.
  \item \textbf{node2vec} \cite{grover2016node2vec} extends DeepWalk by breadth-first and depth-first walk strategies.
  \item \textbf{SDNE} \cite{wang2016structural} simultaneously preserve rectified first-order and second-order proximity with the help of deep learning.
  \item \textbf{AROPE} \cite{zhang2018arbitrary} preserves vanilla arbitrary-order proximity via efficient matrix factorization. %Let AROPE($k$) denote the AROPE that preserves vanilla proximity up to $k^{th}$ order.
\end{itemize}

For all methods except GraRep, the dimension of learned embeddings is set to 128. The remaining unspecified parameters are set to the values recommended by paper authors or manually finetuned to the best.

Similar to LINE, we have EPINE(1st), EPINE(2nd) and EPINE(1st+2nd). For all experiments, we set $\lambda_i=0.1^{i-2}$ and $k=3$ for BlogCatalog, $k=2$ for others. It is not easy to select the best $\eta$ automatically, but $\eta$ $\in \{17, 11, 5\}\times 10^{-4}$ always yields best performance.\footnote{We use $\eta = 17\times 10^{-4}$ for BlogCatalog and Flickr, $\{11, 5\}\times 10^{-4}$ for Wikipedia and YouTube respectively.} $\eta$ and number of training samples are positively related to the density and edge number of networks respectively. Most of the rest of parameters are the same as LINE. 

\begin{table}
\begin{center}
\caption{The AUC sores of network reconstruction and link prediction on BlogCatalog}
\begin{tabular}{|c|c|c|}
\hline
Method & \tabincell{c}{Network\\Reconstruction} & \tabincell{c}{Link\\Prediction} \\
\hline
DeepWalk & 0.9513 & 0.9434 \\
LINE & 0.9511 & 0.9491 \\
GraRep & 0.9555 & \textbf{0.9556}\\
node2vec & 0.9467 & 0.9420 \\
SDNE & 0.9510 & 0.9484 \\
AROPE & 0.9488 & 0.9437\\
\hline
EPINE(1st) & 0.7743 & 0.7745 \\
EPINE(2nd) & 0.9605 & 0.9547 \\
EPINE(1st+2nd) & \textbf{0.9609} & \textbf{0.9556} \\
\hline
\end{tabular}
\label{tab2}
\end{center}
\end{table}

\begin{table*}
\begin{center}
\caption{The results of node classification on various datasets}
\begin{tabular}{|c|c|c|c|c|c|c|c|c|}
\hline
\multirow{2}{*}{Method} & \multicolumn{2}{c|}{Wikipedia} & \multicolumn{2}{c|}{BlogCatalog} & \multicolumn{2}{c|}{Flickr} & \multicolumn{2}{c|}{YouTube}\\
\cline{2-9}
& Micro-F1 & Macro-F1 & Micro-F1 & Macro-F1 & Micro-F1 & Macro-F1 & Micro-F1 & Macro-F1 \\
\hline
DeepWalk & 0.5030 & 0.1030 & 0.4274 & 0.2865 & 0.4216 & 0.3035 & 0.3028 & 0.2115\\
LINE & 0.5831 & 0.1731 & 0.4327 & 0.2919 & 0.4273 & 0.3162 & 0.3086 & \textbf{0.2285}\\
GraRep & 0.5422 & 0.1244 & 0.4265 & 0.2875 & --- & --- & --- & ---\\
node2vec & 0.5488 & 0.1243 & 0.4133 & 0.2769 & 0.4119 & 0.2872 & 0.3083 & 0.2209\\
SDNE & 0.4351 & 0.0649 & 0.3047 & 0.1319 & 0.3480 & 0.2054 & --- & ---\\
AROPE & 0.5439 & 0.1488 & 0.3391 & 0.1723 & 0.3136 & 0.1587 & --- & ---\\
\hline
EPINE(1st) & 0.5465 & 0.1224 & \textbf{0.4467} & 0.3068 & 0.4162 & 0.2914 & 0.3021 & 0.2016\\
EPINE(2nd) & \textbf{0.5971} & \textbf{0.1863} & 0.4396 & 0.3087 & 0.4273 & 0.3071 & \textbf{0.3106} & 0.2223\\
EPINE(1st+2nd) & 0.5950 & 0.1773 & 0.4450 & \textbf{0.3143} & \textbf{0.4350} & \textbf{0.3237} & 0.3089 & 0.2262\\
\hline
\end{tabular}
\label{tab3}
\end{center}
\end{table*}

\subsection{Network Topological Information Preserving}

Tasks of network reconstruction and link prediction evaluate if node embeddings preserve the network topological structure information, which is the most basic goal of NE. As in \cite{shi2019network}, we represent each edge by concatenating the embeddings of two endpoints. 

For network reconstruction, we randomly sample 80\% connected edges and the same number of unconnected edges to train the LIBLINEAR classifiers. The rest of connected edges and the same number of unconnected edges are utilized as test samples. 

For link prediction, we randomly remove 40\% of the edges without breaking the connectivity of the network. After network representation learning, we use the existing edges and the same number of originally unconnected edges as training samples, the removed links and the same number of originally unconnected links as test samples.

The results on BlogCatalog are reported in Table \ref{tab2}. EPINE outperforms others in network reconstruction and reaches comparable performance with GraRep in link prediction. Compared with GraRep, EPINE is more scalable.

\subsection{Network Semantic Information Preserving}

The task node classification is to predict the node categories based on node representations. It evaluates to what extent node representations preserve the high-level semantic information of networks. In experiments, node embeddings are fed directly into the LIBLINEAR classifiers. We use 90\% labeled nodes as training samples, and 10\% as test ones. The results are reported in table \ref{tab3}.\footnote{We exclude some of the baseline results due to efficiency problems or memory errors. The server we use has two Intel(R) Xeon(R) CPU E5-2630 v4 @ 2.20GHz and 320G memory.} All the values are the average of several runs for the sake of result stability.\footnote{500, 250, 20, 50 runs for Wikipedia, BlogCatalog, Flickr and YouTube respectively.} EPINE achieves comparable performance on YouTube and outperforms all baselines on others. 

\subsection{Different Orders of Rectified Proximity}
\begin{figure}[tbp]
\centering
\includegraphics[height=1.6in]{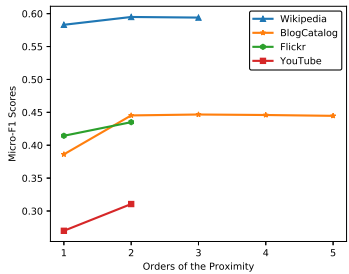}
\caption{Effects of different order proximity.}
\label{fig4}
\end{figure}

We set different $k$ for node classification and record the results in Figure \ref{fig4}. The $A^{4-order}$ and $A^{6-order}$ for Wikipedia and BlogCatalog are both all zeros. The $A^{3-order}$s for Flickr and YouTube are too dense to efficiently learn node embeddings. In spite of this, we can see that only second order rectified proximity brings significant improvements, hence we could simply set $k=2$ in practice.

\begin{table}
\begin{center}
\caption{Effects of certain EPINE calculation steps.}
\begin{tabular}{|c|c|c|c|}
\hline
 & Calculation Step & \tabincell{c}{Wikipedia} & \tabincell{c}{BlogCatalog} \\
\hline
 & LINE & 0.5831 & 0.4327\\
1 & + reweighting & --- & 0.3861 \\
2 & + rectified second-order & 0.5852 & 0.4358 \\
2a & w/o reweighting & --- & 0.4200 \\
3 & + add-dot & 0.5942 & 0.4359 \\
3a & rectified $\to$ vanilla & 0.5827 & 0.4416 \\
4 & + truncating (EPINE) & 0.5950 & 0.4453 \\
\hline

\end{tabular}
\label{tab3}
\end{center}
\end{table}

\subsection{Ablilation Studies}
Take LINE as base, we construct EPINE step by step, and report Micro-F1 results of the node classification at each step in Table \ref{tab3}. Row 2a and 3a does not belong to the construction. Reweighting (row 1) or rectified second-order proximity alone (row 2a) would degenerate the performance, but once we combine them (row 2), the Micro-F1 would slightly exceed LINE. Then we substitute normal matrix multiplication by the additive one, evident improvement occurs on Wikipedia (comparing row 2 and 3). At this step, if we remove masking, that is, substitute rectified proximity by the vanilla one (row 3a), performance will get worse on Wikipedia, but become better on BlogCatalog. This is because BlogCatalog is a social network and vanilla second-order proximity strengthen the importance of edges that form triangular structures (see Figure \ref{fig2.sub.1}), which is a special case.

\section{Conclusions}\label{sec5}
In this work, we propose EPINE, a novel approach that further exploits the information carried by adjacency matrices. To be specific, EPINE provides a feasible way for preserving edge weight information into node embeddings, and a scalable way to accurately calculate the high-order proximity, which allows studying the effect of specific $k$-order proximity. Comprehensive experiments demonstrate the effectiveness of our method. Enhanced proximity information makes improvements.

In the future, we will focus on searching better methods for edge reweighting and EPINE similarity preserving.

\section*{Acknowledgments}
We are grateful to Chongjun Wang for his fruitful comments and advice.

\newpage
\bibliographystyle{named}
\bibliography{epine}

\end{document}